\documentclass[prb,twocolumn,showpacs,preprintnumbers,amsmath,amssymb, superscriptaddress]{revtex4}

\usepackage{graphicx}
\usepackage{dcolumn} 

\begin{document}

\title{Implicit solvation model for density-functional study of nanocrystal surfaces and reaction pathways}

\author{Kiran Mathew}
\email{km468@cornell.edu}
\affiliation{Department of Materials Science, Cornell University, Ithaca, New York 14853, USA}
\author{Ravishankar Sundararaman}
\affiliation{Department of Physics, Cornell University, Ithaca, New York 14853, USA}
\author{Kendra Letchworth-Weaver}
\affiliation{Department of Physics, Cornell University, Ithaca, New York 14853, USA}
\author{T. A. Arias}
\affiliation{Department of Physics, Cornell University, Ithaca, New York 14853, USA}
\author{Richard G. Hennig}
\email{rhennig@cornell.edu}
\affiliation{Department of Materials Science, Cornell University, Ithaca, New York 14853, USA}

\date{\today}

\begin{abstract}
Solid-liquid interfaces are at the heart of many modern-day technologies and provide a challenge to many materials simulation methods. A realistic first-principles computational study of such systems entails the inclusion of solvent effects. In this work we implement an implicit solvation model that has a firm theoretical foundation into the widely used density-functional code VASP. The implicit solvation model follows the framework of joint density functional theory.  We describe the framework, our algorithm and implementation, and benchmarks for small molecular systems. We apply the solvation model to study the surface energies of different facets of semiconducting and metallic nanocrystals and the S$_{\text{N}} 2$ reaction pathway. We find that solvation reduces the surface energies of the nanocrystals, especially for the semiconducting ones and increases the energy barrier of the S$_{\text{N}} 2$ reaction.
\end{abstract}


\maketitle

\section{\label{sec:1}Introduction}
Recently, scientists have determined that the understanding of nanoparticle synthesis and electrochemical interfaces is crucial to designing novel materials for energy technology and improving the performance characteristics of batteries and catalysts.\cite{choi,bealing12, robinson, abruna} The physics of solid-liquid interfaces plays a major role in the synthesis of nanoparticles, chemical reactions at electrode surfaces,\cite{npelectrochem} and in many other phenomena important to energy applications. The thermodynamics and kinetics of nanoparticle interfaces determine the particle morphologies and surface states, which in turn affect the self-assembly as well as optical and electronic properties of these materials.\cite{hanrath, murray, frankwise}

A comprehensive understanding of nanoparticle synthesis and electrochemical interfaces via experiments presents a challenge due to the heterogeneity of the interface and the complexities of the solid and liquid materials involved.\cite{npinterface} Computational studies provide an alternative method to improve our fundamental understanding of solid/liquid interfaces and to predict properties of novel materials interfaces.\cite{bealing12, ta4, ta6, gygi}

There are two main ways to achieve a computational treatment of solid/liquid interfaces. If a complete {\it ab-initio} treatment of the solute/solvent system is desired, all solvent molecules must be considered explicitly. Thus, to reach the equilibrium state we need to relax the electronic and ionic degrees of freedom of both the solute and the solvent molecules. This treatment is quite expensive, as the number of solvent molecules in the system required to capture the essential equilibrium properties is huge and because of the statistical averaging required for the solvent molecules.

An alternative approach is to treat the solute quantum-mechanically and to treat the solvent as a continuum, which means that the solute is immersed in a bath of solvent and the average over the solvent degrees of freedom becomes implicit in the properties of the solvent bath. 
Implicit solvation models for plane wave density-functional theory (DFT) codes were pioneered by Fattebert and Gygi,\cite{gygi} independently developed and placed into the rigorous framework of joint density functional theory (JDFT) by Arias {\it et al.},\cite{ta5} and extended by Marzari {\it et al.} to include a model for cavitation and dispersion.\cite{Marzari}  These methods provide a much more computationally tractable way to vary the electronic and the geometric degrees of freedom of the solute so that the ground state of the combined solute/solvent system conforms with the equilibrium properties of the solvent bath. Since the solute electronic structure is still being treated quantum-mechanically, this approach can be quite accurate assuming all interactions between the solute and the solvent are considered in proper detail.

For polar or ionic solute systems in contact with polar fluids, the electrostatic interaction between the solute and the solvent is the most significant solvation effect. For nonpolar solutes and solvents, the van der Waals interaction can dominate over electrostatics. For large molecules, the energy required to form a cavity in the solvent is the most important contribution to the solvation energy.  Thus, any solvation theory which can be generally applicable to nanoparticles, molecules, and surfaces must consider all of these effects. In this work, we review an implicit solvation model which places a quantum-mechanical solute in a cavity surrounded by a continuum dielectric description of the solvent. Describing the dielectric response as a functional of the solute electronic charge density leads to a self-consistent determination of the cavity by considering the polarization of the solvent by the electronic structure of solute, the effects of cavitation and dispersion, and the corresponding response of the solute system to the presence of the solvent. This implicit solvation model provides a computationally efficient and accurate technique for understanding solute/solvent interfaces.  
 

Following the approach of Refs.~[\onlinecite{ta4}] and [\onlinecite{ta5}], we briefly review the theoretical underpinnings and framework of these implicit solvation models in Sec.~\ref{sec:framework}.  We then describe in Sec.~\ref{sec:implementation} the implementation of an implicit solvation model derived from joint density functional theory\cite{ta4,ta5,ta11} in VASP, a widely used and multi-featured plane-wave DFT code. Though this model was previously implemented in the open source codes DFT++\cite{ta2} and JDFTx,\cite{ta9} implementation in VASP places this theory into the self-consistent field framework for the first time. Due to the plane-wave basis, this implementation is more scalable for large periodic systems than solvation models which employ Gaussian type orbitals.  Additional advantages of this new implementation include higher performance, better MPI parallel scaling, and the interoperability with an extensive library of standardized ultrasoft pseudopotential and projector-augmented wave potentials. In Sec.~\ref{sec:validation} we benchmark the accuracy of this implementation by calculating molecular solvation energies, and comparing against both experimental and JDFTx-calculated values.
Finally, in Section~\ref{sec:applications} we apply the model to metal and semiconductor nanocrystal interfaces and reaction pathways.  We find that the implicit solvation model that we have implemented into VASP provides an efficient and accurate approach to determine solvation energies of molecular and extended systems. Also, the solvation modifications are freely available as a patch to the original VASP source code.\cite{vaspsol}

\section{Theoretical framework of implicit solvation model}
\label{sec:framework}

Following Refs.~[\onlinecite{ta5}] and [\onlinecite{ta10}], the free energy, $A$, of the combined solute/solvent system can be written as a sum of two terms, a universal functional $F$ of the total electron density and the thermodynamically averaged atomic densities of the solvent species, and a term describing the electrostatic energy contribution
\begin{eqnarray} \label{eq:0}
A &=& F[n_\mathrm{tot},\{N_{i}( \vec r)\}] + \nonumber \\
&& +\int d^3r\, V_\mathrm{ext}(\vec r) \left (\sum_i Z_i N_i(\vec r) - n_\mathrm{tot}(\vec r) \right ).
\end{eqnarray}
Here $n_\mathrm{tot}(\vec r)$ is the total electron density, which is the sum of the electron density of the solute and the solvent, i.e. $n_\mathrm{tot}(\vec r)  = n_\mathrm{solute}(\vec r) + n_\mathrm{solv}(\vec r)$. $N_i(\vec r)$ are the thermodynamically averaged atomic densities associated with the chemical species $i$ in the solvent, $V_\mathrm{ext}(\vec r)$ is the external potential due to the solute nuclei and $F$ is a universal functional. The functional $F$ is universal in the sense that it depends only on the electron density and the solvent atomic densities.

Next we use the variational principle for the thermodynamic free energy of the electron-nuclear system in a fixed external electrostatic potential $V_\mathrm{ext}(\vec r)$ to determine the ground-state free energy of the system
\begin{eqnarray} \label{eq:1}
A_0 &=& \min_{n_\mathrm{tot},\{N_i(\vec r)\}} \Bigg \{ F[n_\mathrm{tot},\{N_{i}(\vec r)\}]  \nonumber \\ 
&& +  \int d^3r\, V_\mathrm{ext}(\vec r) \left (\sum_i Z_i N_i(\vec r) - n_\mathrm{tot}(\vec r) \right) \Bigg \}.
\end{eqnarray}

Although the above formalism provides an exact DFT treatment of the combined solute/solvent system, it is difficult to solve in practice due to the minimization involved over the immense number of solvent degrees of freedom. In order to make it amenable to a computational treatment, the free energy is first minimized over the solvent electron density and then over the solute electron density to determine the ground state free energy.\cite{ta5}

Minimizing Eq.~\eqref{eq:0} with respect to the solvent electron density $n_\mathrm{solv}$, we obtain
\begin{eqnarray}  \label{eq:3}
\tilde A &=& G[n_\mathrm{solute}(\vec r),\{N_i(\vec r)\}, V_\mathrm{ext}(\vec r)] \nonumber \\
&& - \int d^3r\, V_\mathrm{ext}(\vec r) n_\mathrm{solute}(\vec r),
\end{eqnarray}
 where
\begin{align}  \label{eq:4}
&G[n_\mathrm{solute}(\vec r),\{N_i(\vec r)\}, V_\mathrm{ext}(\vec r)] = \min_{n_\mathrm{solv}}  \Bigg \{F[n_\mathrm{tot},\{N_i(\vec r)\}] \nonumber \\
& \qquad {}- \int d^3r\, V_\mathrm{ext}(\vec r)  \left (\sum_i Z_i N_i(\vec r) - n_\mathrm{solv}(\vec r) \right)  \Bigg \}.
\end{align}
$G$ is a universal functional of the electron density of the solute $n_\mathrm{solute}(\vec r)$, the average atomic densities of the various species in the solvent $\{N_i(\vec r)\}$, and the external potential of the solute nuclei $V_\mathrm{ext}(\vec r)$.  The functional $G$ can be separated as following
 \begin{align}  \label{eq:5}
& G[n_\mathrm{solute}(\vec r),\{N_{i}(\vec r)\},V_\mathrm{ext}(\vec r)]  =  A_\mathrm{KS}[n_\mathrm{solute}(\vec r),V_\mathrm{ext}(\vec r)]  + \nonumber \\
&\qquad {} + A_\mathrm{diel}[n_\mathrm{solute}(\vec r),\{N_{i}(\vec r)\},V_\mathrm{ext}(\vec r)],
\end{align}
where $A_\mathrm{KS}$ is the usual Kohn-Sham density functional for the solute and $A_\mathrm{diel} $ is the term that encapsulates all the interactions of the solute with the solvent and the internal energy of the solvent. To further simplify the expression, the functional $A_\mathrm{diel} $ is minimized with respect to the average atomic densities of the solvent, $N_i(\vec r)$,
\begin{align}  \label{eq:6}
&\tilde{A}_\mathrm{diel}[n_\mathrm{solute}(\vec r),V_\mathrm{ext}(\vec r)] = \nonumber \\
& \qquad {} \min_{\{N_i(\vec r)\}} A_\mathrm{diel}[n_\mathrm{solute}(\vec r),\{N_{i}(\vec r)\},V_\mathrm{ext}(\vec r)].
\end{align}
Combining Eqs.~\eqref{eq:1} to \eqref{eq:6} leads to the ground state free energy of the solute/solvent system,
\begin{align}  \label{eq:8}
&A_0  = \min_{n_\mathrm{solute}(\vec r)} \bigg \{A_\mathrm{KS}[n_\mathrm{solute}(\vec r),V_\mathrm{ext}(\vec r)] \\ \nonumber 
& \qquad {} -  \int d^3r\, V_\mathrm{ext}(\vec r) n_\mathrm{solute}(\vec r)+  \tilde{A}_\mathrm{diel}[n_\mathrm{solute}(\vec r),V_\mathrm{ext}(\vec r)] \bigg \}.
\end{align}

Importantly, this minimization procedure leads to a free energy of the combined solute-solvent system written as a functional of only the electron density of the solute, $n_\mathrm{solute}(\vec r)$, and the external potential of the solute nuclei, $V_\mathrm{ext}(\vec r)$, properties determined solely by the solute. All the solvent effects are contained in the functional $\tilde{A}_\mathrm{diel}$, which is obtained by the minimization over the solvent electron density and the thermodynamically average atomic densities of the solvent. Thus, the functional $\tilde{A}_\mathrm{diel}$ describes a continuum model for the solvent, which has an equilibrium structure fully determined by the properties of the solute, upon the solute electronic structure. 
The minimization of the functionals in Eq.~\eqref{eq:8} with respect to the solute degrees of freedom leads to the ground state free energy of the joint system. Up to this point, the theory is exact, although the exact form of $\tilde{A}_\mathrm{diel}$ is unknown. Approximations must be made to the functional $\tilde{A}_\mathrm{diel}$ for practical calculations.

As a first approximation, we consider the electrostatic interaction between the solute and the solvent, which affects the equilibrium polarization of the solvent dipoles. Assuming that the solvent polarization depends linearly on the electric field for the range of fields encountered in the vicinity of the solute, the solvent polarization can be described by the local relative permittivity of the solvent, $\epsilon(\vec r)$. We must then include in the functional $\tilde{A}_\mathrm{diel}$ a term to account for the electrostatic interaction between the solute electronic structure and the corresponding bound charge distribution induced in the solvent.\cite{gygi,ta5}


However, an electrostatic-only approach is insufficient to describe solvation of molecules and nanoparticles, where cavitation and dispersion may play a significant role.  Since the non-electrostatic effects are concentrated in the first solvation shell, to describe these effects,\cite{cramer} we adopt a version of the empirical model proposed by Marzari {\it et al.},\cite{Marzari} and placed into the joint density-functional theory framework by Arias {\it et al.}\cite{ta11} that describes the corrections as an interface term that is proportional to the solvent-accessible area. 
Thus, we also include in $\tilde{A}_\mathrm{diel}$ an additional term to describe the free energy contributions of cavitation and dispersion, 
\begin{align}  \label{eq:13}
A_\mathrm{cav} = \tau \int dr |\nabla S|,
\end{align}
where $\tau$ is the effective surface tension parameter, which describes the cavitation, dispersion and the repulsion interaction between the solute and the solvent that are not captured by the electrostatic terms alone
 and $S(\vec r)$ is the cavity shape function described below.

Decoupling the electrostatic term from the Kohn-Sham functional $A_\mathrm{KS}$ and combining it with the interaction term and the cavitation term, we obtain
\begin{align}  \label{eq: 9}
A[n_\mathrm{solute}(\vec r), \phi(\vec r)]  = & A_\mathrm{TXC}[n_\mathrm{solute}(\vec r)] \nonumber \\
& + \int d^3r\, \phi (\vec r)\left ( N_\mathrm{solute}(\vec r) - n_\mathrm{solute}(\vec r) \right ) \nonumber \\
& -  \int d^3r\, \epsilon(\vec r) \frac{| \nabla \phi |^2}{8 \pi} \nonumber \\
& + A_\mathrm{cav} ,
\end{align}
where $A_\mathrm{TXC}$ is the free energy density functional describing the kinetic and exchange-correlation energy of the solute and $N_\mathrm{solute}(\vec r)$ is the solute nuclear charge density.

We make an important distinction between the potentials $\phi(\vec r)$ and $V_\mathrm{ext}(\vec r)$.  $\phi(\vec r)$ is the combined electrostatic potential due to the electronic ($n_\mathrm{solute}(\vec r)$) and nuclear ($N_\mathrm{solute}(\vec r)$) charges of the solute system in a polarizable medium. $V_\mathrm{ext}(\vec r)$ is the potential due to the nuclei in the solute, and is usually described by pseudopotentials or the projector-augmented wave method.  Outside a specified cutoff radius it has the form $\frac{Z_\mathrm{eff}}{r}$, where $Z_\mathrm{eff}$ is the effective charge of the respective atom. Since the solvent described by $\epsilon(\vec r)$ does not penetrate the core region of the pseudopotentials, we can approximate the contribution of the nuclear charges to the combined electrostatic potential $\phi(\vec r)$ of the solute by a sum over terms of the form $\frac{Z_\mathrm{eff}}{r}$.

So far, we have described a solute system surrounded by a dielectric medium quantified by the relative permittivity of the solvent system, $\epsilon(\vec r)$. However, we must also determine the form of the dielectric cavity formed in the solvent by the solute. Implicit solvation models offen differ in their approximations for this cavity. A common way to construct the cavity is to place spheres around the solute atoms and then take the union of these overlapping spheres.\cite{PCM-Review} Inside the so-formed cavity the relative permittivity is assumed to be that of vacuum, outside it takes the value of the solvent, and the induced charges are placed on the surface of this cavity. One might also assume a diffuse dielectric cavity such that the relative permittivity changes continuously.

\begin{figure}[t]
  \includegraphics[width=7.5cm]{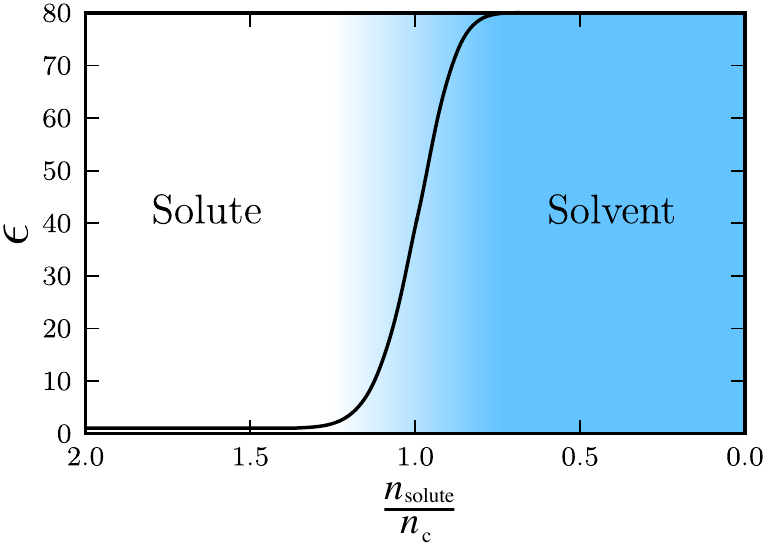}
  \caption{\label{fig:eps_n} Smooth variation of the relative permittivity $\epsilon$ from the vacuum value of one to the value of the solvent, e.g. 80 for water.}
\end{figure}

In this work, we assume a diffuse dielectric cavity that is a local functional of the electronic charge density of the solute, i.e for the relative permittivity $\epsilon(\vec r) = \epsilon(n_\mathrm{solute}(\vec r))$. This assumption leads to a diffuse cavity that is implicitly determined by the electronic structure of the solute. The smooth transition into the cavity also ensures that the derivatives of the energy functional are continuous, thereby simplifying the implementation of the geometric optimization of the solute system. We assume the following functional dependence\cite{ta5} of the relative permittivity of the solvent on the solute electronic charge density
\begin{equation}  \label{eq: 10}
\epsilon(n_\mathrm{solute}(\vec r)) = 1 + (\epsilon_\mathrm{b} - 1) S(n_\mathrm{solute}(\vec r)),
\end{equation}
where $\epsilon_\mathrm{b}$ is the relative permittivity of the bulk solvent and $S(n_\mathrm{solute}(\vec r))$ is the cavity shape function, given by
\begin{equation}  \label{eq: 14}
S(n_\mathrm{solute}(\vec r)) = \frac{1}{2}\text{erfc} \left \{ \frac{\text{log}(n_\mathrm{solute}/n_\mathrm{c})}{\sigma \sqrt{2}} \right \}.
\end{equation}
The parameter $n_\mathrm{c}$ determines at what value of the electron density the dielectric cavity forms, and $\sigma$ is the parameter that determines the width of the diffuse cavity.
Figure~\ref{fig:eps_n} illustrates the dependence of the permittivity on the solute electronic charge density. The above functional form of the relative permittivity ensures that the value of the relative permittivity varies smoothly from one in the bulk of the solute to $\epsilon_\mathrm{b}$ in the bulk of the solvent. This gradual variation emulates the first solvation shell effects, i.e the value of the relative permittivity of the solvent close to the solute is smaller than the equilibrium bulk value due to the higher electric field near the solute surface, a phenomenon known as dielectric saturation.

As shown by Refs.~[\onlinecite{ta4}]~and~[\onlinecite{ta5}], the functional in Eq.~\eqref{eq: 9} can be optimized by first minimizing with respect to the electrostatic potential, $\phi(\vec r)$, and then with respect to the solute electronic charge density, $n_\mathrm{solute}(\vec r)$. Minimization with respect to $\phi(\vec r)$ leads to a generalized Poisson equation\cite{gygi}
\begin{align}  \label{eq: 11}
& \nabla \cdot \left [ \epsilon (n_\mathrm{solute}(\vec r)) \nabla \phi \right ] = \nonumber \\
& \qquad  -4 \pi \left \{ N_\mathrm{solute}(\vec r) - n_\mathrm{solute}(\vec r) \right \},
\end{align}
where $N_\mathrm{solute}(\vec r)$ consists of the effective core charges approximated by Gaussians as described below and $n_\mathrm{solute}(\vec r)$ is the valence electronic charge density. Minimization of Eq.~\eqref{eq: 9} with respect to the electronic charge density, $n_\mathrm{solute}(\vec r)$, yields the typical Kohn-Sham Hamiltonian with the following additional terms in the local part of the potential
\begin{equation}  \label{eq: 12}
V_\mathrm{solv} = \frac{d \epsilon(n_\mathrm{solute}(\vec r))}{d n_\mathrm{solute}(\vec r)} \frac{| \nabla \phi |^2}{8 \pi} + \tau \frac{d |\nabla S|}{d n_\mathrm{solute}(\vec r)}.
\end{equation}

Corrections to the Hellman-Feynman force terms should also be made due the modifications of the Kohn-Sham potential. The force corrections consist of two terms,
\begin{align}  \label{eq: 13}
  \int \Delta \phi \frac{d N_\mathrm{solute}}{d R_I}  d^3r + \int V_\mathrm{solv} \frac{d n_\mathrm{solute}}{d R_I} d^3r .
\end{align}
The first term is due to the change in the electrostatic potential $\Delta \phi$, which is the difference between the solution to Eq.~\eqref{eq: 11} and the electrostatic potential when the relative permittivity is one.
The second term is due to the augmentation of the electronic charge density with the pseudo-charge density at the atom locations to prevent the fluid from entering in the core region, as described in the following section.

\section{Implementation of implicit solvation model}
\label{sec:implementation}

The implicit solvation model reviewed above has been implemented into the Vienna Ab-initio Software Package (VASP),\cite{vasp} a widely-used plane-wave DFT code. Combined with VASP's parallel scalability to large system sizes and the availability of established and tested libraries of ultrasoft pseudo-potentials (USPP)\cite{uspp,kresseuspp} and projector-augmented wave (PAW) potentials,\cite{paw} the MPI compatible implementation extends the capabilities of the software to study large solvated metallic and semiconducting systems in an efficient manner. 

The VASP code solves the Kohn-Sham equations through self-consistent iterations to find the electronic ground state. To include the description of the solvent effects, we modify the local potential of the Kohn-Sham Hamiltonian and the expression for the total free energy. The solution to the generalized Poisson equation, given by Eq.~\eqref{eq: 11}, must become part of the self consistent loop as the valence charge density changes in each self-consistent iteration.

\begin{figure}[t]
  \includegraphics[width=8.5cm]{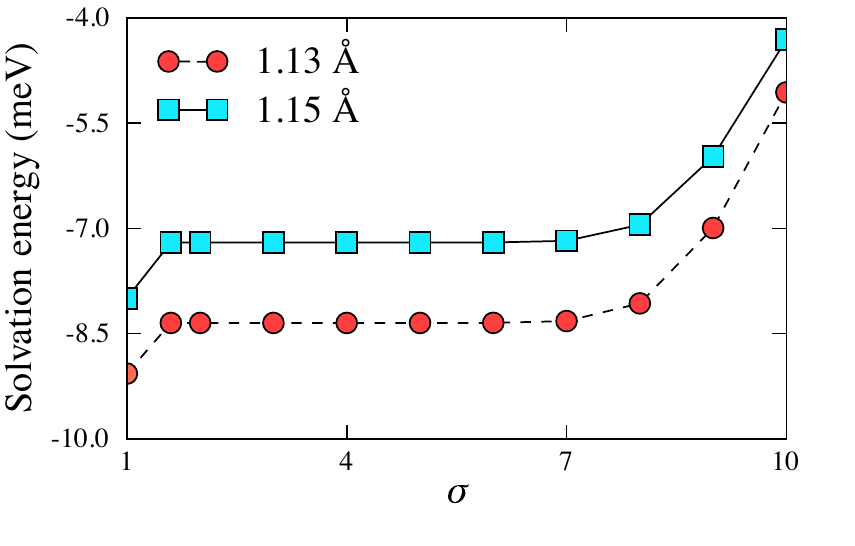}
  \caption{\label{fig:e_s} Solvation energy as function of the Gaussian ionic width for CO molecule for two different atomic separations, where the Gaussian width, $\sigma$, is specified in units of the grid size.}
\end{figure}

The generalized Poisson equation is solved in each electronic step to obtain the electrostatic potential of the combined solute electronic charge density and ionic charge density in the polarizable medium that describes the solvent. Since VASP is a plane-wave DFT code, we take advantage of fast Fourier transformation (FFT) methods and approximate the nuclear point charges by Gaussians of finite width,\cite{ta5}
\begin{equation}  \label{eq: 12}
N_\mathrm{solute}(\vec r) = \sum_I \frac{Z_I}{ (2\pi \sigma)^{3/2}} \exp \left({-\frac{(\vec r - \vec R_I)^2}{2 \sigma^2}}\right),
\end{equation}
where $\sigma$ is the width of the Gaussian, $\vec R_I$ and $Z_I$ are the positions and charges, respectively, of the nuclear species $I$ in the solute. As long as the width, $\sigma$ is sufficiently small such that the Gaussians do not interfere with the solvent, the interaction energy does not depend on the Gaussian width. Figure~\ref{fig:e_s} illustrates for the CO molecule that the solvation energy is independent of the choice of $\sigma$ over a range of values of $\sigma$. For small values of $\sigma$, i.e. when $\sigma$ is of the order of the grid spacing or smaller, deviations occur because of the reduced numerical accuracy of the integration of the Gaussians. For large values of $\sigma$, the deviations are caused by the Gaussians reaching into the region of the solvent.

For DFT implementations that use pseudopotentials such as VASP,\cite{vasp} PWSCF,\cite{pwscf} ABINIT,\cite{abinit} JDFTx,\cite{ta9} etc., the electronic charge density corresponds only to the valence charge density which tapers off close to the atomic cores. Since the solvent effects described by the permittivity are assumed to be a functional of the local electronic charge density, the possibility exists that, due to the reduced valence charge density near the cores, the relative permittivity could become greater than one in the region of the atomic cores. Following Ref.~[\onlinecite{ta4}], in order to ward off such unphysical solvent penetration into the atomic core regions, pseudo-charges centered at the atomic cores are added to the valence charge density. This is strictly a numerical device and has no effect on the interaction energies as these pseudo charges are only used in the computation of the relative permittivity of the solvent. In principle, one could also replace these pseudo-charges with the partial core charges from an appropriately chosen pseudopotential, as is done in the current JDFTx implementation of the same fluid model.\cite{ta9}

The generalized Poisson equation, Eq.~\eqref{eq: 11}, is solved using a pre-conditioned conjugate gradient algorithm. We use a pre-conditioner of the form $\frac{1}{G^2}$, where $G$ is the nonzero reciprocal lattice vector. This choice of pre-conditioner gives the exact solution to the Poisson equation in Fourier space when the permittivity is constant. 

The solution procedure using the conjugate gradient algorithm is made efficient through the use of FFTs for the evaluation of the gradient and divergence terms in Eq.~\eqref{eq: 11}. First, the gradient term, $\nabla \phi$, is evaluated in Fourier space. It is then transformed to real space and multiplied with the spatially varying permittivity, $\epsilon(\vec r)$, which is given as a functional of the charge density. Then the divergence of the term $\epsilon(\vec r) \nabla \phi$ is computed in the Fourier space using a FFT, leading to the complete Fourier space representation of the right hand side of Eq.~\eqref{eq: 11}.


\begin{figure}[t]
  \includegraphics[width=8.5cm]{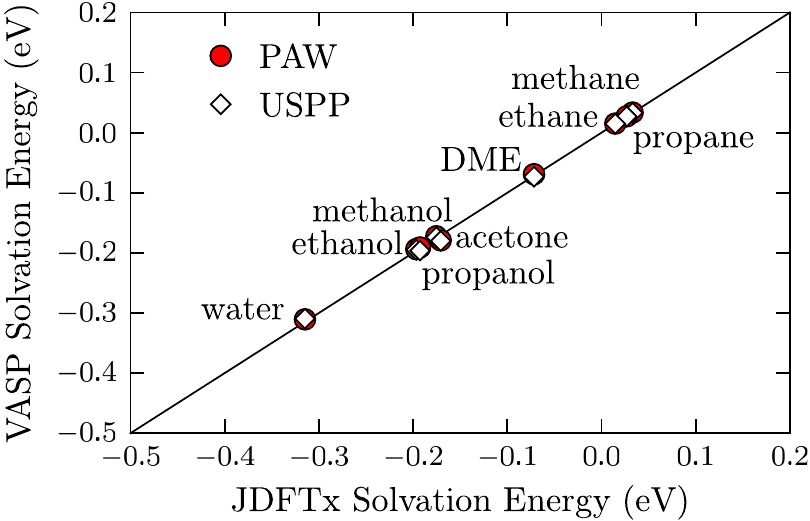}
  \caption{\label{fig:jdft_vasp} Comparison of solvation energies for different molecules in water calculated with VASP and JDFTx. The JDFTx calculations employed norm-conserving pseudopotentials and the VASP calculations ultrasoft pseudopotentials and the PAW method.}
\end{figure}

For the shape function parameters $n_c$ and $\sigma$ in  Eq.~\eqref{eq: 14} and the effective surface tension $\tau$ in Eq.~\eqref{eq:13} we use the values of Ref.~[\onlinecite{ta11}]  that were obtained by a fit of the model to experimental solvation energies for molecules in water. The specific values are $n_c = 0.0025 \text{\AA}^{-3}$, $\sigma = 0.6$, and $\tau = 0.525 \text{meV/\AA}^2$.

The implementation of this solvation model in VASP is parallelized over multiple processors using MPI. To demonstrate the efficiency of the implementation, we calculate the surface energy of a Pt (111) surface slab with 5 layers of Pt and a 10~\AA\ slab spacing. The vacuum calculation on 64 cores converged in 39 seconds requiring 28 self-consistent iterations and the same system solvated in water, starting from the vacuum wave functions, converged in 35 seconds requiring 16 self-consistent iterations.

\section{Validation of implicit solvation model}
\label{sec:validation}

We validate the correct implementation of our solvation model for the energies and forces by comparing the solvation energies of several molecules with values obtained for the same solvation model from the JDFTx code\cite{ta9} and with experimental results. For the forces, we compare the values from the implemented analytic expressions with the values obtained by numerical differentiation of the energy.  

The calculations for the validation in this section as well as the applications in Sec.~\ref{sec:applications} are performed with the modified VASP code using USPP and the PAW method, the PBE exchange-correlation functional. For the molecular systems we use a cutoff energy of 800~eV, the $\Gamma$ point for $k$-point sampling, and a cubic box of 10~\AA\ edge length. The atomic positions are obtained from the computational chemistry comparison and benchmark database.\cite{nist} The calculations of the surface energies use a cutoff energy of 460~eV, a $16 \times 16 \times 1$ mesh for $k$-point sampling, and a vacuum spacing of 10~\AA. The geometry of the platinum slabs for different crystal facets are taken from our previous study given in Ref.~[\onlinecite{Fishman12}].  For PbS, 100 surface geometry had 5 layers, 110 had 10 layers and 111 had 9 layers(reconstructed surface with Pb termination). Fully relaxed vacuum slab geometries were used for both Pt and PbS solvation calculations. For the solvation model parameters, we use the default values of our implementation as specified in the previous section and the relative permittivity of water $\epsilon_b$ = 80, which is the default solvent in the implementation.


\begin{figure}[t]
\includegraphics[width=8.5cm]{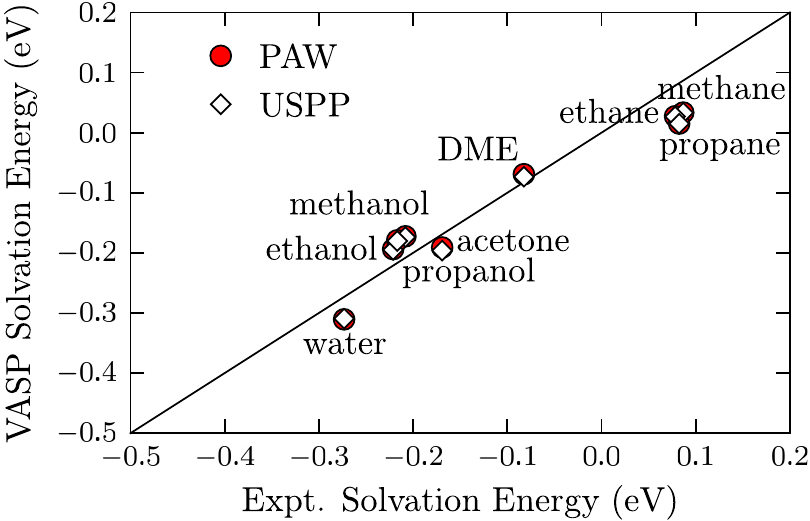}
\caption{\label{fig:exp_vasp} Experimental versus VASP calculated solvation energies for different molecules in water}
\end{figure}

The correctness of the VASP implementation of the solvation model is verified by comparing the solvation energies of organic molecules with the values obtained from the identical solvent model implemented in the JDFTx code. Figure~\ref{fig:jdft_vasp} shows that the solvation energies of both codes are nearly identical, indicating that the method is implemented correctly. The small discrepancies between the VASP and JDFTx solvation energies (on the order of 10~meV) can be attributed to differences in the pseudopotentials used and are well within common pseudopotential errors.\cite{Parker11} 
Figure~\ref{fig:exp_vasp} compares the calculated solvation energies to experimental data. The calculated solvation energies are in good agreement with the experimental values. The experimental and computed solvation energies obtained using JDFTx and VASP are also listed in Table~\ref{tab:3}.


\begin{table}[t]
  \caption{\label{tab:3} Molecular solvation energies; VASP, JDFTx and experimental values. All energies are in eV}
  \begin{ruledtabular}
    \begin{tabular}{ldddd}
      Molecules & \multicolumn{1}{c}{$E_\text{expt}$} & \multicolumn{1}{c}{$E_\text{jdftx}$}
      & \multicolumn{1}{c}{$E_\text{VASP}^\text{PAW}$} & \multicolumn{1}{c}{$E_\text{VASP}^\text{USPP}$}  \\
      \colrule
      Acetone & -0.17 & -0.19 & -0.19 & -0.20 \\
      Dimethyl ether & -0.08 & -0.07 & -0.07 & -0.07 \\
      Ethane & +0.08 & +0.03 & +0.03 & +0.03\\
      Ethanol & -0.22 & -0.17 & -0.18 & -0.18\\
      Methane & +0.08 & +0.01 & +0.02 & +0.01\\
      Methanol & -0.22 & -0.20 & -0.19 &-0.19 \\
      Propane & +0.09 & +0.03 & +0.03 & +0.03\\
      Propanol & -0.21 & -0.18 & -0.17 & -0.17\\
      Water & -0.27 & -0.31 & -0.31 &  -0.31\\
    \end{tabular}
  \end{ruledtabular}
\end{table}


The implicit solvation model results in correction terms for the forces, which are derived from the energy expression of Eq.~\eqref{eq: 9} and given by Eq.~\eqref{eq: 13}. We confirmed the numerical accuracy of the forces by comparing the results of the implemented analytic expressions with the values obtained by numerical differentiation of the energy for several molecules.



 
\section{Applications}
\label{sec:applications}
We apply the implemented solvation model to surfaces of materials that are of current technological interest and study the effect of various solvents on the surface energies of the dominant facets of metal and semiconductor nanocrystals. In addition, we determine the energy barrier for the nucleophilic substitution reaction of chloromethane and compare the results with quantum chemistry calculations.

\subsection{Solvation effects on metal and semiconductor nanocrystals}

Optical, electronic, and magnetic properties of nanocrystals strongly depend on their size and shape. These properties are in turn affected by the functional groups present on the surface and the type of solvent in which they are dispersed. Here we consider platinum and lead sulphide nanocrystals to ascertain how the presence of a solvent affects the surface energies of different nanocrystal facets and what the implications are for the nanocrystal shape. Platinum nanocrystals have a wide range of applications in catalysis from fuel cells to catalytic converters.\cite{pt1,pt2} Lead sulphide nanocrystals have exceptional optical properties\cite{pbs1} and are considered as emerging novel materials for inorganic-organic bulk hybrid solar cells\cite{pbs2} and tunable near infrared detectors.\cite{pbs3}

\begin{figure}[t]
  \includegraphics[width=8cm]{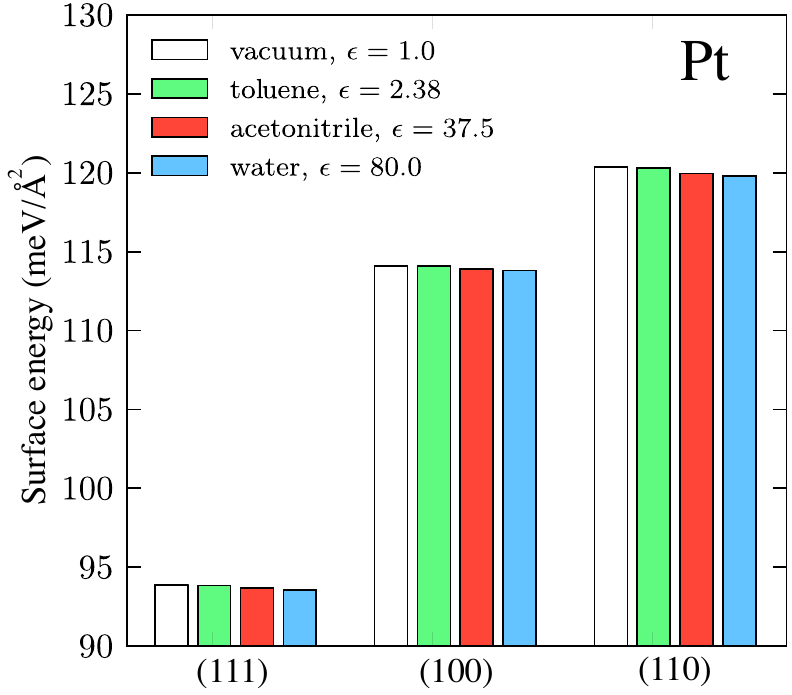}
  \caption{\label{fig:pt} Surface energies of the (111), (100), and (110) facets of Pt nanocrystals in different solvents.}
\end{figure}

\begin{figure}[t]
  \includegraphics[width=8cm]{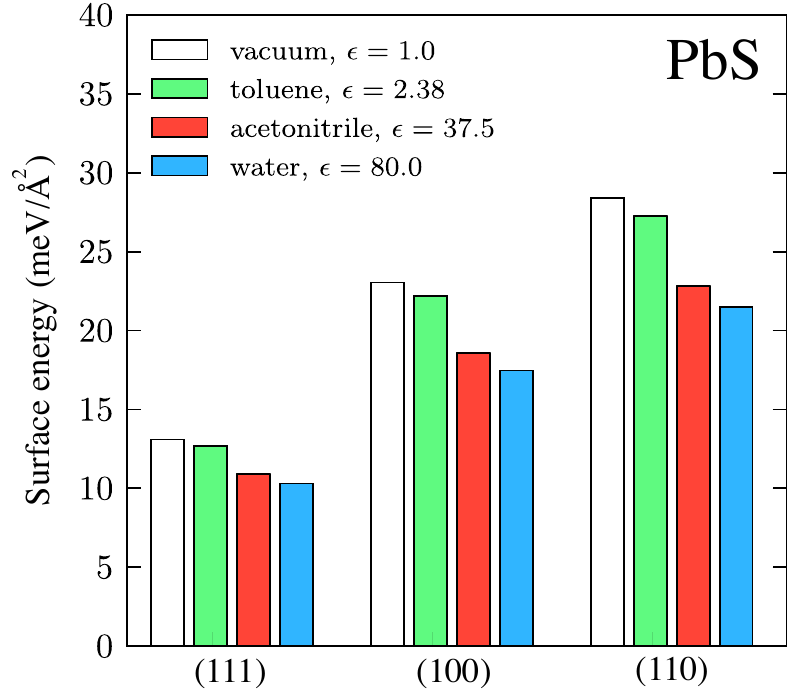}
  \caption{\label{fig:pbs} Surface energies of the (100), (111), and (110) facets of PbS nanocrystals in different solvents.}
\end{figure}

Figures~\ref{fig:pt} and~\ref{fig:pbs} show how the presence of solvent affects the surface energies of the low-energy facets of Pt and PbS nanocrystals. In all cases, the solvent reduces the surface energies with the more polar solvents resulting in higher reductions. The reduction in surface energies for Pt are up to 2~meV/\AA$^2$ and for PbS up to 7~meV/\AA$^2$.

The more significant effect of solvation on the PbS surfaces than the Pt surfaces is due to the nature of bonding in these systems. PbS exhibits a partially ionic bonding, while the Pt bonding is purely metallic. The presence of partially ionic bonds on the PbS surfaces lead to stronger electric fields at the surface experiencing solvent screening.
Due to the reasonably high electric fields present at the PbS surfaces, we also confirm that a linear dielectric response to electric field strength is sufficient for capturing the solvation energy of these materials. For the PbS surfaces, the surface energy differences between the linear and nonlinear model calculated with the JDFTx code\cite{ta11} is less than 2\%. Ref.~[\onlinecite{ta11}] has confirmed that the effect of nonlinearity is also negligible for metal surfaces. 

We observe that the facets with higher surface energies are more stabilized by solvation than the surfaces with lower energies, an effect which is particularly noticeable for the PbS surfaces. This phenomenon leads to a more isotropic surface energy distribution in the presence of polar solvent than in vacuum, which can affect the nanocrystal morphology.\cite{bealing12}  We predict that the presence of polar solvent results in more spherical and less strongly faceted PbS nanocrystals.

\subsection{Reaction pathways}

Reaction pathways and barriers are also influenced by the presence of solvents.\cite{rxn1,rxn2} To demonstrate the importance of solvation effects in determining the reaction pathways and to illustrate the capability of the current implementation, we consider the nucleophilic substitution S$_\text{N}$2 reaction of Cl$^-$ and CH$_3$Cl. This bimolecular nucleophilic substitution plays an important role in physical organic chemistry and hydration increases the reaction energy barrier, which increases the the transfer rate by 20 orders of magnitude.\cite{ch3clcl}

\begin{figure}[t]

  \includegraphics[width=8.5cm]{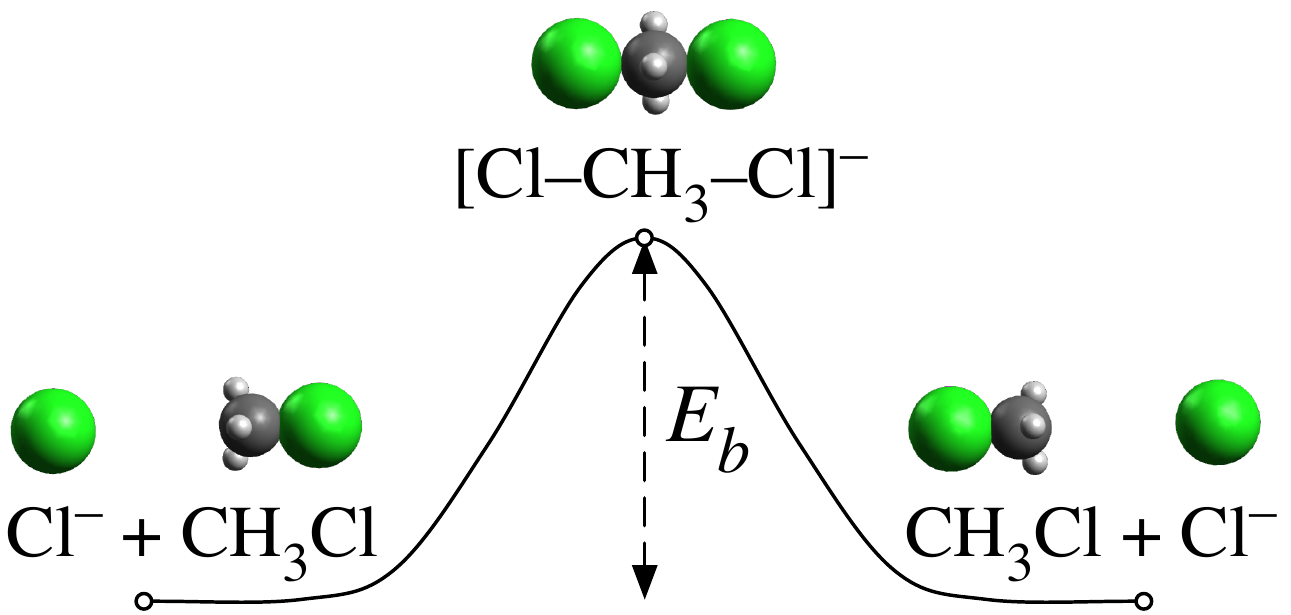}
  \caption{\label{fig:rxn_pth} Nucleophilic substitution S$_\text{N}$2 reaction of a chlorine ion with chloromethane.}
\end{figure}

\begin{table}[t]
  \caption{\label{tab:2} Energy barriers for the nucleophilic substitution S$_\text{N}$2 in vacuum and in water calculated with VASP, Gaussian09, and constrained {\it ab-initio} molecular dynamics.}
  \begin{ruledtabular}
    \begin{tabular}{lcc}
     & $E_b^\text{vacuum}$ (eV) & $E_b^\text{water}$ (eV)\\
    \hline
    VASP &0.34 &0.60\\
    Fattebert and Gygi\footnote[1]{Ref.~[\onlinecite{gygi}]} & & 0.61\\
    Gaussian 09\footnote[2]{Cavity of atom-centered spheres.}  & 0.32 &0.63  \\ 
    Gaussian 09\footnote[3]{Static isodensity model of Ref.~[\onlinecite{Foresman96}].}  & 0.32 &0.69  \\
    {\it ab-initio} molecular dynamics\footnote[4]{Ref.~[\onlinecite{ensing}]} & & 0.82\\
   \end{tabular}
 \end{ruledtabular}
\end{table}

Figure~\ref{fig:rxn_pth} illustrates the pathway for the $\text{S}_\text{N}\text{2}$ reaction Cl$^-$ + CH$_3$Cl $\rightleftharpoons$ ClCH$_3$ + Cl$^-$, where $E_b$ is the energy barrier. We calculate the energy barriers for this reaction in vacuum and water using VASP and Gaussian09.\cite{g09} The VASP calculations employ a cubic box with 25~\AA\ edge length, a cutoff energy of 800~eV, and our implemented solvation model. The Gaussian09 calculations use the aug-cc-pV5Z basis set and the static isodensity solvation model,\cite{Foresman96}  which is similar to the solvation model we implemented in VASP.


Table~\ref{tab:2} compares the energy barrier for the S$_\text{N}$2 reaction obtained with VASP with various other methods. We find that the energy barriers obtained from VASP and Gaussian09 are in good agreement. We also observe that the energy barrier in Gaussian09 only weakly depends on the solvation model. The energy barriers obtained with our solvation model in VASP and Gaussian09 also compare well with the result of Fattebert and Gygi\cite{gygi} of 0.61~eV which neglected the contribution from the cavitation.\cite{gygi} For the case of this reaction energy barrier, neglecting the cavitation energy is a good approximation since the cavity does not change much during the reaction. The reaction barrier obtained with constrained {\it ab-initio} molecular dynamics simulations with explicit solvent is 0.82~eV\cite{ensing}, about 0.2~eV higher than the values for the implicit solvation model. This difference may be due to anharmonic contributions to the energy barrier.




\section{\label{sec:4}Conclusions}

We implemented an implicit solvation model that describes the effect of electrostatics,\cite{ta5} cavitation, and dispersion\cite{Marzari} on the interaction between a solute and solvent into the plane-wave DFT code VASP. The model was validated by comparing the values from the VASP implementation with the values from the JDFTx implementation and experimental data. Our implementation provides a computationally efficient means to calculate the effects of solvation on molecules and crystal surfaces. We apply the solvation model to determine the effects of solvation on the different facets of metal and semiconductor nanocrystals and the energy barrier for the nucleophilic substitution reaction of chloromethane. Solvation significantly reduces the surface energies of the semiconducting PbS nanocrystals and only weakly affects the surface energies of the metallic Pt nanocrystals. For the nucleophilic substitution reaction we obtain energy barriers in good agreement with previous calculations. The strength of our solvation model implementation is its capability to handle large periodic systems such as metal and semiconductor surfaces and its interoperability with standard ultrasoft pseudopotential and projector-augmented wave potential libraries. The software is freely available as a patch to the original VASP code.\cite{vaspsol}


\begin{acknowledgments}
This work was supported by the Energy Materials Center at Cornell (EMC2) funded by the U.S. Department of Energy, Office of Science, Office of Basic Energy Sciences under award number DE-SC0001086 and by the National Science Foundation under award number CAREER DMR-1056587. This research used computational resources of the Texas Advanced Computing Center under Contract Number TG-DMR050028N.
\end{acknowledgments}

\providecommand{\noopsort}[1]{}\providecommand{\singleletter}[1]{#1}%

\end{document}